\documentclass{IEEEtran}
\usepackage{cite}
\usepackage{amsmath,amssymb,amsfonts}
\usepackage{graphicx}
\usepackage{textcomp,nicefrac}
\usepackage{subcaption}
\usepackage{multirow}
\usepackage[numbers]{natbib}
\usepackage{xcolor}
\usepackage{hyperref}
\hypersetup{
    colorlinks=true,
    citecolor={teal},
    linkcolor={teal},
}

\def\BibTeX{{\rm B\kern-.05em{\sc i\kern-.025em b}\kern-.08em
T\kern-.1667em\lower.7ex\hbox{E}\kern-.125emX}}
\begin{document}
\title{In-orbit Spectral Calibration Prospects for the COSI Space Telescope}
    
    
    
    

\author{
    \IEEEauthorblockN{Aravind B. Valluvan, Steven E. Boggs, Savitri Gallego, Jarred Roberts, Gabriel Brewster, Sophia Haight, Carolyn Kierans, Sean Pike, Albert Y. Shih, John A. Tomsick, Andreas Zoglauer}
    
\thanks{A. B. Valluvan, S. E. Boggs, J. Roberts, G. Brewster, S. Haight, and S. Pike are with the Department of Astronomy and Astrophysics, UC San Diego, CA 92093 USA}
\thanks{S. Gallego is with Johannes Gutenberg-Universit\"at Mainz, 55122 Mainz, Germany}
\thanks{C. Kierans and A. Y. Shih are with NASA Goddard Space Flight Center, MD 20771 USA}
\thanks{J. A. Tomsick and A. Zoglauer are with the Space Sciences Laboratory, UC Berkeley, CA 94720 USA}
\thanks{Corresponding author: A.B. Valluvan (e-mail: avalluvan@ucsd.edu)}
}

\maketitle

\begin{abstract}
The Compton Spectrometer and Imager is an upcoming NASA space telescope in the MeV range. COSI’s primary science goals include precisely mapping positron annihilation and nuclear line emissions in the Milky Way galaxy through Compton imaging. This relies on our ability to maintain COSI’s spectral performance over its mission lifetime. Changes to the detectors' gain characteristics over time will result in inaccurate measurement of astrophysical gamma-ray energies. Moreover, observations from past MeV telescopes and proton-beam experiments have shown that radiation damage in space causes photopeak shifts and spectral line broadening. These necessitate a plan for regular, in-orbit spectral calibration. In this study, we demonstrate a method to monitor and recalibrate the COSI detectors using background line emissions produced by the space radiation environment. We employ Monte Carlo simulations of particle background and show that strong background lines arise from nuclear excitation of COSI's detectors (germanium) and cryostat (aluminum) materials. These span COSI's entire bandwidth for single-site interactions and can be used to monitor the effects of radiation damage and gain shifts every twelve hours at the full instrument level and every 16 days at the individual detector level. Methods developed by Pike et al. \citep{pike2023trapping,pike2025holetrappingproton} to correct the effects of hole trapping and gain characteristics can then be applied to recover the original spectral performance.
These results inform COSI’s telemetry requirements for calibration and housekeeping data, and rule out the need for an on board radioactive calibration source which would have increased the complexity of the spacecraft. 

\end{abstract}


\section{Introduction}
\label{sec:introduction}
The Compton Spectrometer and Imager (COSI) is an upcoming NASA Small Explorer satellite mission, with a planned launch in August 2027 to a low Earth, equatorial orbit. COSI is a gamma-ray telescope, consisting of 16 high-purity germanium detectors (GeD), and is designed to survey the entire sky from 0.2 to 5 MeV with high spectral resolution \citep{tomsick2023compton}. COSI’s primary science goals include mapping the 511 keV line emission from positron annihilation and gamma-ray line emissions from the decay of radioactive isotopes. Radioactive isotopes are created during shell burning and core-collapse of massive stars, and radioactively-powered transients such as supernovae and binary neutron star mergers. These science goals rely on the ability to maintain COSI’s spectral performance over its mission lifetime, which is at least two years. 

Active efforts are currently directed to maximize the spectral performance of the detectors on the ground. These include collecting data using the GeDs from radioactive sources with known gamma-ray line energies \citep{kierans2015calibration, beechert2022calibrations}, measuring and calibrating the amplitude of the electronic readout signal with physical energy \citep{beechert2022calibrations,sleator2019benchmarking}, and characterizing the effects of differences in electron and hole propagation on spectral line profile \citep{pike2023trapping,boggs2023numerical}. However, exposure to space radiation can affect the characteristics of the instrument \citep{jean2003spi,diehl2018integral,miller2020planning}, which necessitates a plan for regular monitoring and recalibration throughout the mission lifespan once in orbit. The space radiation environment contains high-energy particles that induce hole traps in the GeD, which will offset the photopeak position and broaden the width of measured spectral lines \citep{pike2025holetrappingproton,haight2025proton}. If left uncorrected, such shifts and broadening will be incorrectly attributed to astrophysical source variations.

Furthermore, the detector gain function, which sets the relationship between the electronic readout units (signal voltage) and energy in physical units (keV), can change over time and result in inaccurate measurement of astrophysical gamma-ray energies \citep{diehl2018integral,grefenstette2022gain,lowell2017thesis}. The exact causes are currently under investigation and are beyond the scope of this work. Nevertheless, regular monitoring of multiple spectral lines over COSI's energy range is the only way to monitor changes to the detector gain characteristics once it is placed in orbit. 

Three approaches to in-orbit spectral monitoring and calibration of space telescopes are prevalent in the X-ray and gamma-ray astrophysics literature. {First, cross-calibration observation campaigns of astrophysical sources using multiple telescopes are routinely performed, which can be used as a calibration reference for future telescopes \citep{grant2025international}.} However, unlike observations at energies below 20 keV, there are no persistent, spectrally-diverse, gamma-ray line emitters in the sky \citep{diehl2018integral}.

Second, X-ray missions such as NuSTAR \citep{harrison2013nustar, kitaguchi2014nustarinflight} and AstroSat \citep{singh2014astrosat,vadawale2016astrosatinorbit}
contain a deployable radioactive source on board the spacecraft to illuminate their Cadmium-Zinc-Telluride (CZT) detectors. 
However, unlike the smaller, planar CZT detector configurations in NuSTAR and AstroSat, the large-volume GeDs in COSI cannot be uniformly calibrated with a radioactive source at a single, fixed position. As photon interactions depend on incident angle and depth, effective spectral calibration requires illumination from multiple directions to ensure that the entire GeD volume is properly characterized. This would increase the number of deployable radioactive sources required, and thereby increase the complexity of the spacecraft, the risk of radioactive contamination, and, worryingly, the instrument background rate \citep{rao2010alpha}.

Past MeV gamma-ray missions such as RHESSI \citep{smith2003rhessi} and INTEGRAL/SPI \citep{vedrenne2003spi} have instead relied on the instrument background spectrum created by the space radiation environment to monitor and calibrate their detectors \citep{hajdas2004activationSAAactual,roques2003spiinflightcalibration,weidenspointner2003activationlines,jean2003spi,lonjou2005degradation}. Background gamma rays are created when high-energy particles ($>10$ MeV) that make up the space radiation environment interact and activate (excite) the spacecraft material. 
The excited nuclei later decay and release gamma-rays with precise energies between $10$ keV and $10$ MeV. Numerous gamma ray emission lines dominate the background spectrum, and such a line emission-dominated background is unique to telescopes operating in the MeV range \citep{weidenspointner2003activationlines,hajdas2004activationSAAactual,diehl2018integral}. 

In this study, we present a plan for in-orbit spectral calibration for COSI based on Monte Carlo simulations of this background spectrum. We demonstrate that these results rule out the need for an on board radioactive calibration source, and inform COSI’s telemetry requirements for calibration and housekeeping data.
The paper is outlined as follows: in Section \ref{sec:instrument}, we provide an overview of the COSI instrument followed by a brief description of COSI’s background simulation model in Section \ref{sec:background}. We describe our spectral analysis methods in Section \ref{sec:analysis} where we also define a quantitative metric for activation line strength. We present the results of our analysis in Section \ref{sec:results} followed by a discussion in Section \ref{sec:discussion}. Section \ref{sec:conclusions} contains the conclusions of this study.


\section{COSI Instrument Overview}
\label{sec:instrument}
\begin{figure}
    \centering
    \includegraphics[width=\linewidth]{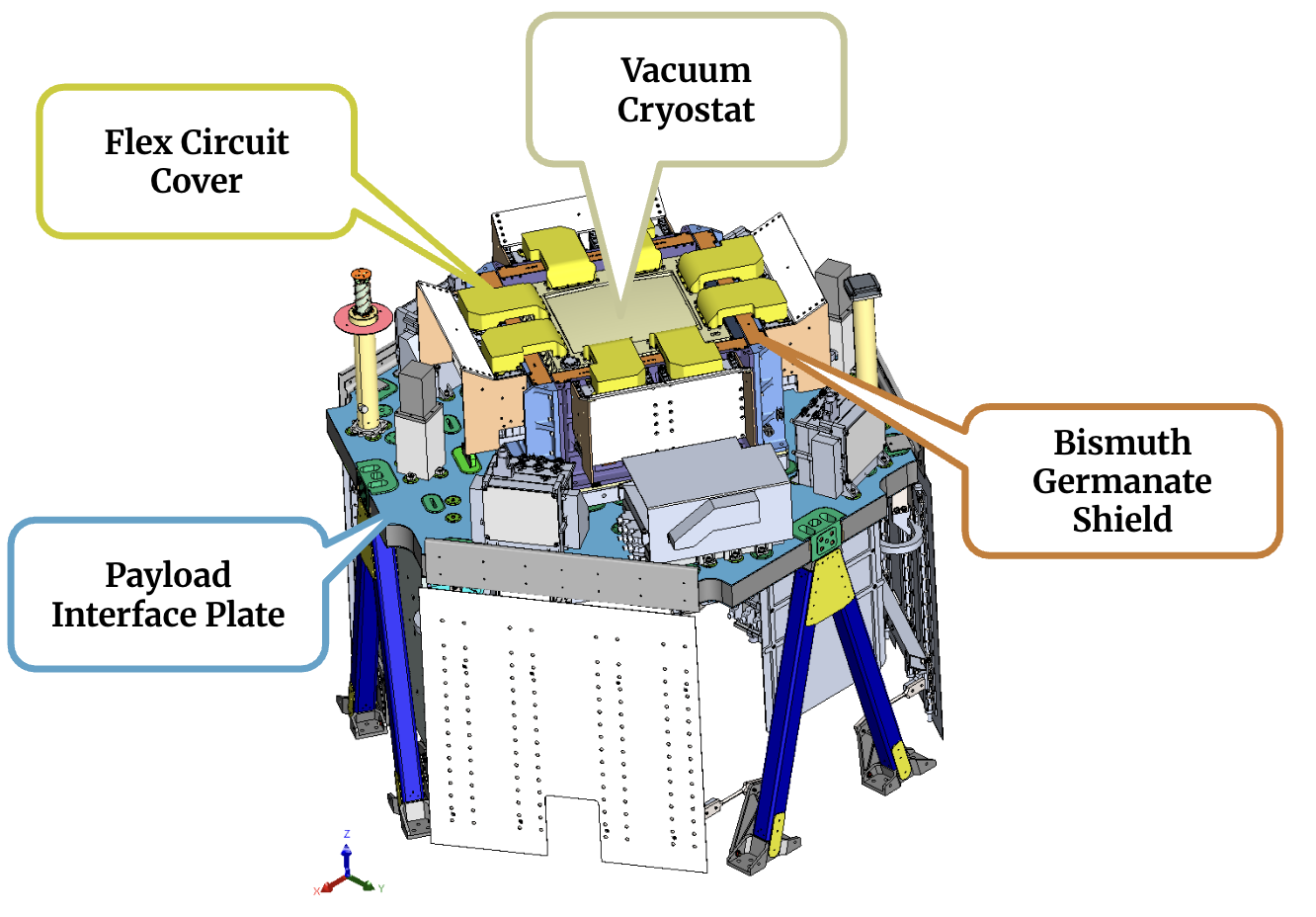}
    \caption{A model of the COSI instrument. The GeD array is enclosed in a vacuum cryostat ({olive-yellow}) and surrounded by bismuth germanate shields ({orange}, see Appendix A). The cryostat and shields are surrounded by eight flex circuits ({bright yellow}), and placed on top of a hexagonal, payload interface plate ({light blue}). The spacecraft bus is not shown here but has been included in our simulations.}
    \label{fig:massmodel}
\end{figure}

The COSI instrument consists of an array of 16 high-purity germanium detectors and is designed to operate in the 0.2--5 MeV energy range. In this energy range, the cross-section for Compton scattering interactions dominates over photoelectric absorption or pair production \citep[Chapter 1]{zoglauer2005thesis}. Thus, unlike X-ray photons which undergo photoelectric absorption, a MeV photon can interact (scatter) multiple times within the instrument before being photoabsorbed. 

As a ``Compact Compton Telescope" (CCT), COSI works by measuring the position (\textbf{r}\textbf{\textsubscript{i}}) and energy deposited (E\textsubscript{i}) at each interaction site \textit{i} for an incident photon. For incident photons with a single interaction site (hereafter referred to as single-site events), the photon energy can be measured but the incident direction cannot be inferred. For incident photons with 2+ interaction sites, the temporal separation between interactions cannot be measured as they are on the order of 100 picoseconds. Instead, 
the most-likely incident direction of the photon is \textit{reconstructed} using the Compton scattering and Klein-Nishina formulae (for a complete overview on the working principle of CCTs, see \citep[Section 3]{kierans2024compton}, \citep[Chapter 4]{zoglauer2005thesis}, and references therein). 

Uncertainties in measuring the energy depositions in CCTs not only impacts spectral performance but will also influence the reconstructed photon direction thereby affecting the angular resolution and overall sensitivity \citep{zoglauer2005thesis, zoglauer2021cosi}. As COSI does not have any coded-mask or optics system to \textit{image} the incoming light, precise individual energy measurements are not only crucial for high-resolution spectroscopic studies but are also essential to create astronomical images. 

\section{Background Simulation Model}
\label{sec:background}
\begin{figure}
    \centering
    \includegraphics[width=\linewidth]{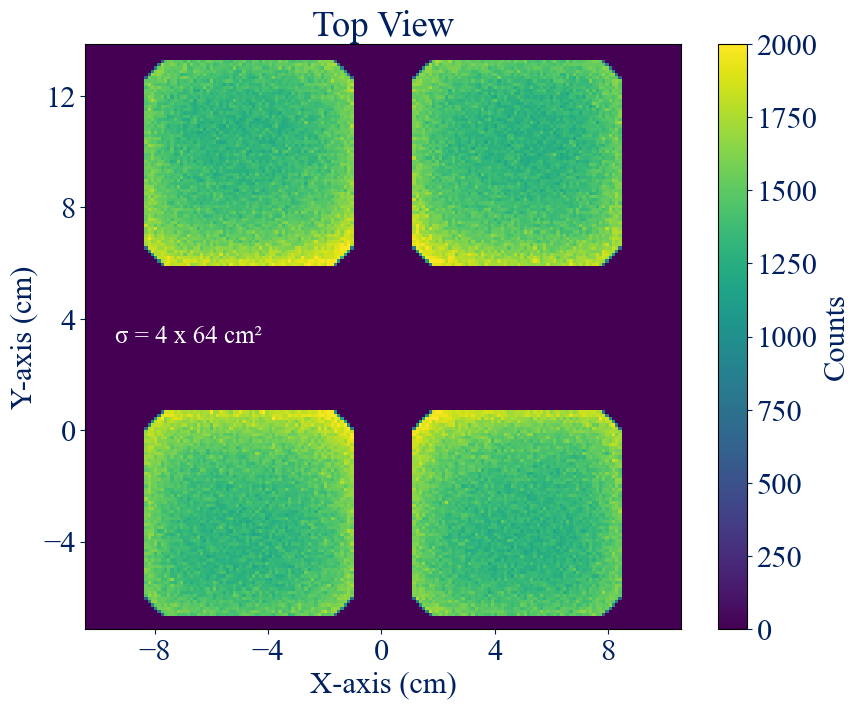}
    \includegraphics[width=\linewidth]{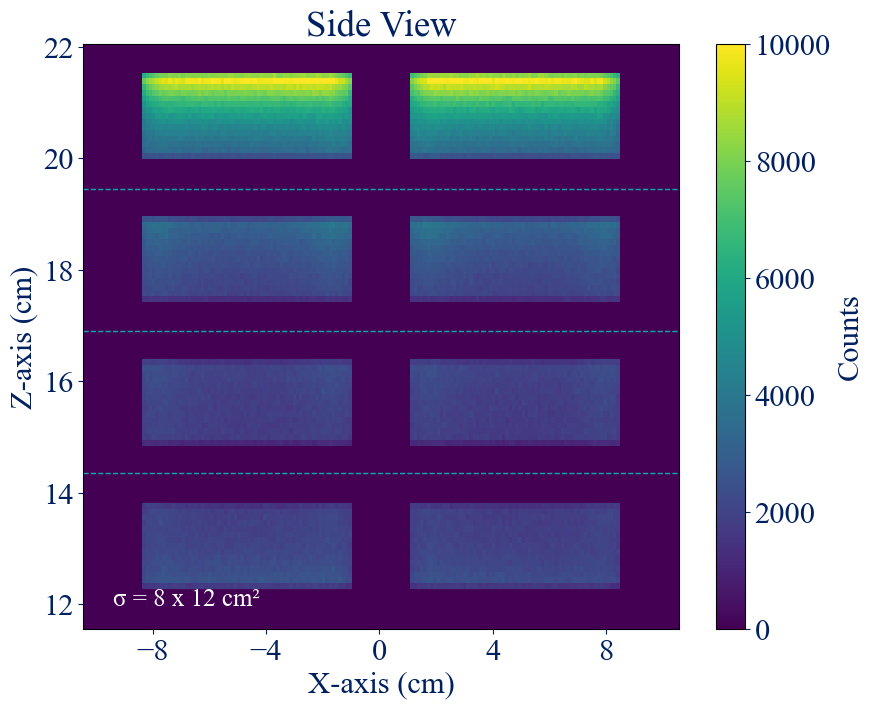}
    \caption{Top (above) and side view (below) of  instrument background for single-site interactions. Each layer in the x-y plane contains four 8 cm $\times$ 8 cm $\times$ 1.5 cm GeDs. The topmost layer is pointed to outer space, and receives the highest photon count rate (primarily continuum background components). The count rates of GeDs at any given layer are uniform within 5\%, and the count rates decrease with depth as fewer photons penetrate that deep. The detector array is enclosed in an aluminum cryostat, which is a source of increased count rates on the edges of the detectors. As this study is focused on monitoring and calibrating the instrument at the level of an individual GeD, all subsequent analyses use data integrated over a GeD module.} 
    \label{fig:sideview}
\end{figure}

Observations at MeV energies are dominated by instrument background induced by activation of detector materials by the space radiation environment. The intensity and diversity of the radiation environment depend not only on cosmic rays but also on solar flare activity, and conditions in the Earth's radiation belts and geomagnetic field. The total radiation is dependent on time and the instrument background varies both on a short, minutes-to-hours timescale as well as with the 11-year solar cycle. COSI has been optimized to maintain a low Earth, equatorial orbit at 530 km (orbital period = 95 minutes) to minimize the impact of the radiation environment, and the overall background levels are predicted to be significantly lower than INTEGRAL/SPI and RHESSI \citep{tomsick2023compton} (for more information, see Appendix A). 

\subsection{Simulation setup}

\textit{Gallego et al.} \citep{gallego2025b} have performed MEGAlib \citep{zoglauer2006megalib} and Geant4-based \citep{agostinelli2003geant4} simulations of particle-matter interactions on COSI. These capture the impact of high-energy neutrons and charged particles over the first three months after launch. The input parameters include COSI’s geometric mass model shown in Fig. \ref{fig:massmodel}, and various energy and spatial distributions for the background particles. Multiple background components have been considered, including cosmic rays, albedo particles reflected off the Earth’s atmosphere, trapped protons encountered during passages through the SAA, and spallated particles arising from primary particle interactions. Long-term variations due to the 11-year solar cycle are yet to be modeled. 

The activation output is postprocessed through the detector effects engine \citep{sleator2019benchmarking,beechert2022calibrations}, a suite of tools used to simulate the readout system and other systematic effects of the COSI detectors, to produce the final background spectrum. \textit{Gallego et al.} \citep{gallego2025bottom} had previously benchmarked the simulation methods and Geant4 versions against observations from the COSI 2016 46-day balloon flight \citep{kierans2017cosiballoon}, a precursor version of the COSI satellite. 
While these simulations incorporate our current best knowledge of the astrophysical input spectra and detector effects engine, we discuss potential sources of systematic uncertainties in Section \ref{subsec:uncertainties}.

\begin{figure*}
    \centering
    \includegraphics[width=\linewidth]{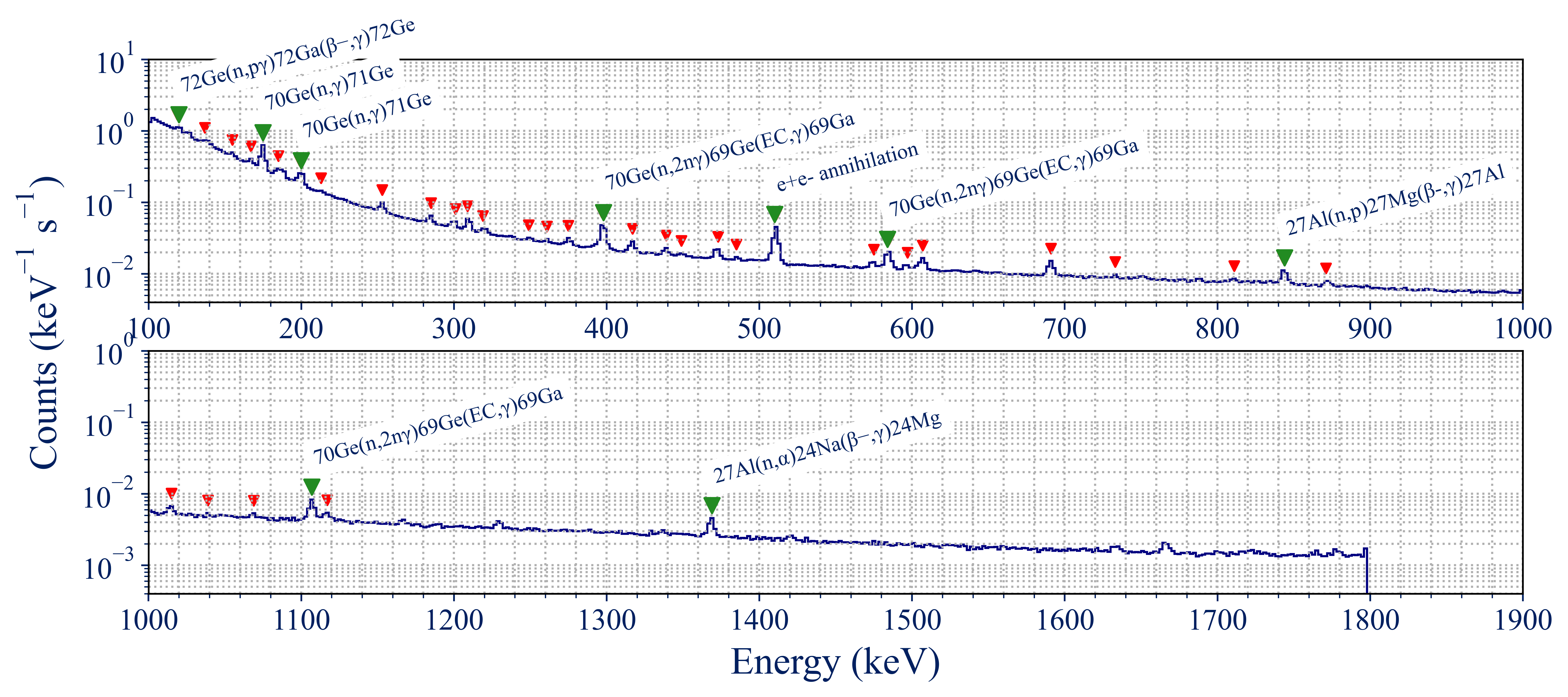}
    \caption{Single-site event rates for the simulated background models. Multiple activation lines are visible and are marked in red. The list of candidate lines selected from this study are marked in green and span the entire energy range. Most of these lines are the result of activating the GeD material or the aluminum cryostat. The spectrum is truncated at 1800 keV -- the saturation energy for single-site interactions used in our simulations.}
    \label{fig:bgsim}
\end{figure*}

\subsection{Data format}

The background simulation outputs a list of photon events with the energy, time and position (x,y,z) of each interaction. Based on the number of interaction sites, each detected photon can be classified as a single-site event (photoabsorption only) or multi-site interaction event (one or more Compton scatters followed by photoabsorption). The GeD readout electronics have sufficient temporal resolution to distinguish between successive incident photons. 

In this study, we only incorporate events that have not been rejected by the anti-coincidence subsystem (see Appendix A). Additionally, we only use single-site events as these provide a direct inference of the energy calibration of the detector without needing an additional step to reconstruct the incident photon direction -- which can introduce additional instrument systematics \citep[Chapter 4]{zoglauer2005thesis}. The spatial distribution of these single-site events is shown in Fig. \ref{fig:sideview}, and the full background spectrum is shown in Fig. \ref{fig:bgsim}. The simulation range for single-site events is truncated at the saturation energy for the readout electronics, which will be at least 1800 keV. 

\subsection{Interpreting the various background components}

The background spectrum can be broadly classified into two components: continuum and line emission \citep{gallego2025b}. The continuum background is dominated by cosmic photons and albedo photons. This component can be effectively modeled as a broken power-law. The line background is dominated by activation due to primary protons, primary alpha particles, as well as SAA protons. Additionally, there are significant contributions to the 511 keV line emission from positrons created in the Earth's atmosphere. Other cosmic-ray and albedo particles are insignificant contributors to COSI’s background. 

The similarities in emitted lines and relative line strengths among the proton, alpha particle, and neutron components indicate that these particles are not directly absorbed. Instead, the high kinetic energy of these particles creates an electromagnetic cascade of neutrons and charged particles with lower kinetic energies that then interact with the payload materials to create activation lines. 
Similar conclusions were drawn by \textit{Jean et al.} and \textit{Diehl et al.} \citep{jean2003spi,diehl2018integral}. This can be attributed to the higher capture cross-sections for lower energy particles compared to higher energy particles. Furthermore, as we will show in Table \ref{table:isotopes}, most activation lines can be traced to neutron captures as opposed to proton or alpha captures.

\section{Analysis Methods}
\label{sec:analysis}
In order to characterize the activation lines that emerge from the Monte Carlo simulations, we first identify strong, isolated activation lines in the simulation and then model individual lines as a Gaussian over a continuum. As multiple lines are visible in Fig. \ref{fig:bgsim}, we 
visually inspected the entire 1800 keV energy range for activation lines that are prominent and not blended with neighboring lines within a $\pm$8 keV energy window (for context, the full-width half-maximum, FWHM, of single-site events as measured in the lab is $\lesssim 6.5$ keV and the wider, 16 keV energy window accounts for potential photopeak offsets.) Selecting such strong, well-isolated lines enables locating them with their relative spacing even when the spectral line profile or instrument gain characteristics change. 

At this stage, 53 activation lines were selected. Within the 16 keV energy windows, we model individual lines as a Gaussian on top of a linear continuum, and calculate their signal-to-noise ratio as \citep{babu1996astrostatistics},

\begin{equation}\label{eq:SNR}
    \text{SNR} = \frac{S}{\sqrt{S+B}} = \frac{st}{\sqrt{st + bt}} \propto \sqrt{t}.
\end{equation}

Here, $t$ is the integration time, $S$ and $B$ are the Gaussian signal and continuum background counts, respectively, within $\pm1.4 \sigma_e$ of the photopeak position, and $\sigma_e = \text{FWHM}/2\sqrt{2\ln2}$ is the instrument's spectral resolution. This energy window maximizes the SNR under a Gaussian model with background \citep{babu1996astrostatistics}. 
The Gaussian signal and continuum background count rates can be approximated as constants $s$ and $b$, respectively. We define the “t\textsubscript{10} metric” as the time taken to reach SNR=10 as a measure for line strength, 
\begin{equation}\label{eq:t10}
    t_{10} = 10^2 \times \frac{s+b}{s^2}.
\end{equation}
The shorter the t\textsubscript{10} value for a line, the better it is for calibration purposes. We derived this metric from the photopeak precision given by \citep{altman2005standard},
\begin{equation}
\delta \mu = \frac{\sigma_e}{\sqrt{S}} \approx \frac{\sigma_e}{\text{SNR}} = \frac{\text{FWHM}}{2\sqrt{2\ln2}\text{ SNR}} \leq 0.05 \text{ FWHM},
\end{equation} 
where the SNR $=\sqrt{S}$ approximation holds for $S\gg B$. Thus, the $t_{10}$ metric will provide sufficiently strong lines to characterize the photopeak position to within 5\% of the FWHM. In Subsection \ref{subsec:impact}, we demonstrate that this precision is sufficient to achieve our science goals. 

To ensure that the temporal variations in the simulation are averaged-out and the model fits are statistically robust, we integrate 7 days of data from the final week of the background simulation model to perform our analyses. This leads to a simpler form for the t\textsubscript{10} metric,
\begin{equation}
    t_{10} = \left(\frac{10}{\text{SNR}(t=7\text{ d})}\right)^2\times 7\text{ d}
\end{equation}

In the next section, we will apply this metric to various activation lines both at the instrument level and at the individual detector level in order to quantitatively determine the strongest lines and study their potential for spectral calibration. We first discuss the expected sources for photopeak shifts. 

\subsection{Effect of Gain Variations}

The detector gain characteristics are expected to change due to variations in detector capacitance, temperature, and performance of readout electronics \citep{lonjou2005degradation, lowell2017thesis}. 
\textit{Lonjou et al.} inferred an energy-dependent correlation between detector temperature and photopeak shift in INTEGRAL/SPI's co-axial GeDs. For temperature fluctuations of $<1$ K, they measured photopeak offsets less than 15 eV at 100 keV and 0.15 keV at 1 MeV. 
While COSI's in-orbit temperature fluctuation timescales and temperature-gain dependence are yet to be characterized, these results inform us of the order of changes to expect.

\subsection{Effect of Radiation Damage}

While the Monte Carlo background simulations output Gaussian-like activation lines whose width is defined by the instrument resolution, radiation damage can affect this line shape \citep{diehl2018integral,pike2023trapping,pike2025holetrappingproton}. Impinging high-energy particles can damage the crystal lattice structure of detectors and create hole traps. These hole traps affect the measured energy, effectively increasing the photopeak line width and inducing a low-energy tail \citep{haight2025proton}. For an estimated high-energy particle fluence of $1.1 \times 10^8$ p\textsuperscript{+}/cm\textsuperscript{2} (95\% confidence level) over COSI's two-year primary mission \citep{haight2025proton, gallego2025b}, we expect photopeak offsets of $\sim 0.5\%$ on the anode side and $\sim 1.5\%$ on the cathode side \citep[Figure 7]{pike2025holetrappingproton}. 

\subsection{Impact of Photopeak Offsets}\label{subsec:impact}

We illustrate and quantify the impact of photopeak offsets on COSI's primary science goals with an example. A 1 keV offset at the galactic \textsuperscript{26}Al decay line 1809 keV translates to a Doppler shift of $\Delta v \sim c\cdot \Delta E/E_0 \sim$ 165 km/s \citep{kretschmer2013kinematics}, which is on the same order as the Doppler shift induced by the Sun's orbital velocity around the galactic center $\sim220$ km/s. Thus, to resolve finer \textsuperscript{26}Al structures and infer the correct velocities, we need to characterize and correct offsets on the order of $\lesssim 1$ keV or $0.05\%$ of $E_0 = 1809$ keV. 
\textit{Haight et al.} and \textit{Pike et al.} inferred a photopeak offset of $\sim 1\% \times E_0$ photopeak offset (18 keV at $E_0=1809$ keV) over the course of two years \citep{haight2025proton,pike2025holetrappingproton}. Thus, we need to recalibrate our instrument on timescales at least 20x faster, i.e., every 36 days, to ensure COSI's photopeak precision is $\lesssim 1$ keV. By using line emissions with SNR$\geq$10, we can attain a photopeak precision of $\lesssim 0.5$ keV at 1809 keV (for COSI's FWHM resolution requirement of $\lesssim 10$ keV \citep{tomsick2023compton}), which is comfortably lower than the required 1 keV precision. 
Hence, as long as the background lines are sufficiently strong for recalibration every 36 days, i.e., $t_{10} < 36$ days, we will have no impact in achieving COSI's science goals. 

\section{Results}
\label{sec:results}
\begin{figure}
    \centering
    \includegraphics[width=\linewidth]{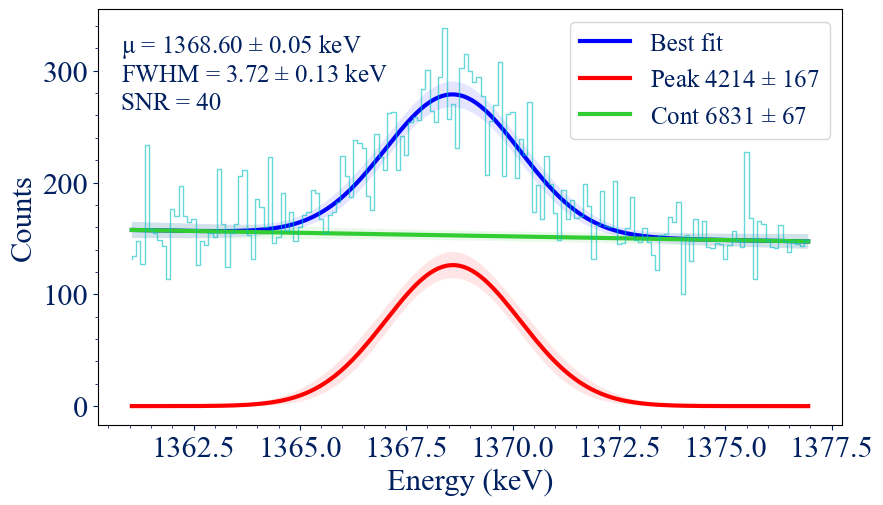}
    \caption{Fitting the 1369 keV, 7-day integrated data with a Gaussian+line model, along with their 99.7\% confidence intervals. The peak and continuum counts on the top right and the SNR value on the top left are calculated using the counts within $\pm 1.4\sigma_e$ of the photopeak. 
    }
    \label{fig:modelfit}
\end{figure}

\begin{figure}
    \centering
    \includegraphics[width=\linewidth]{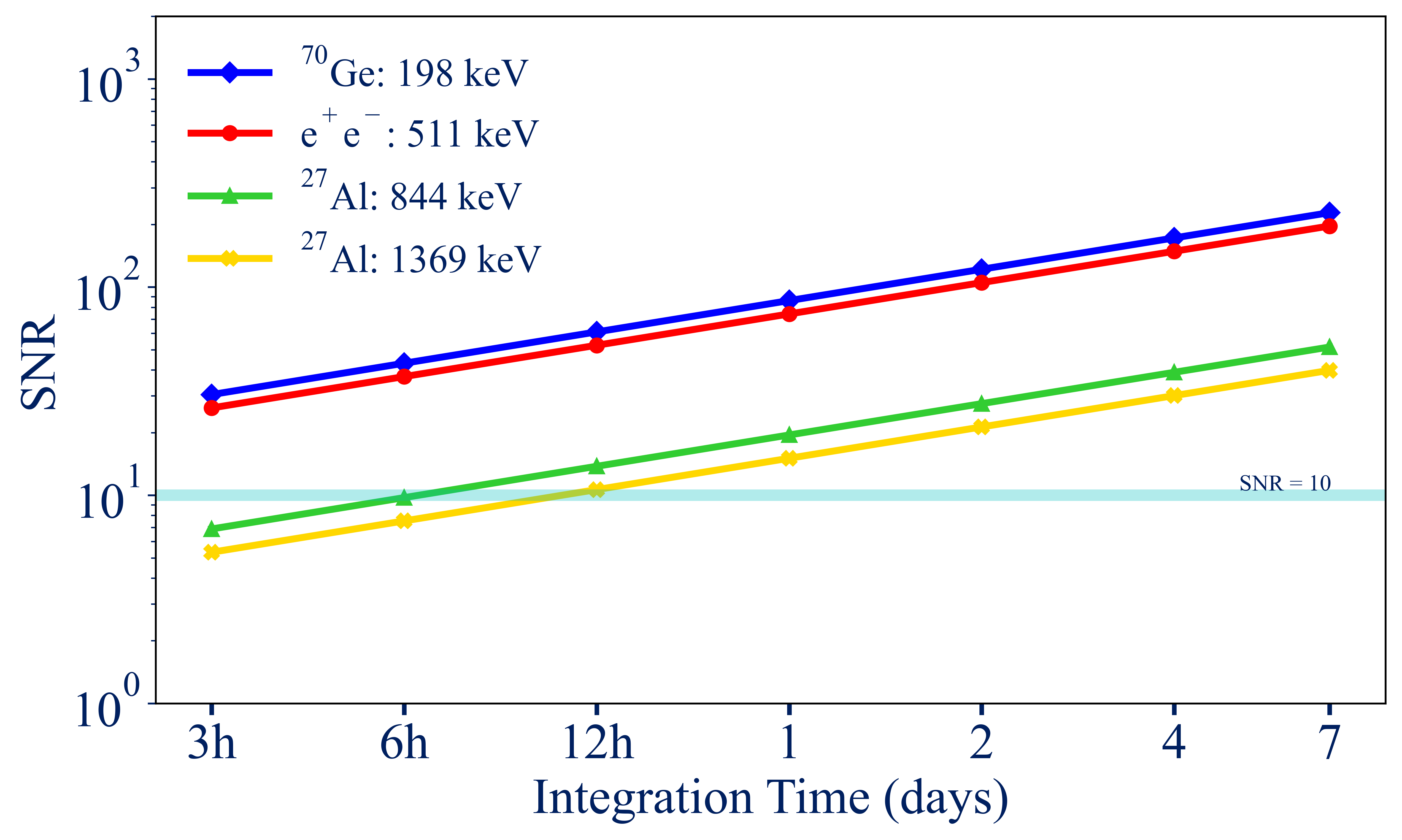}
    \caption{The evolution of the SNR value with time for various activation lines is shown here, with photon data integrated over the entire instrument. The horizontal line denotes SNR=10 and the integration time required to cross this horizontal line is defined as the t\textsubscript{10} value. The slope is $\nicefrac{1}{2}$ as expected from Equation \ref{eq:SNR}.}
    \label{fig:linevsSNR}
\end{figure}

\begin{figure*}
    \centering
    \includegraphics[width=0.49\linewidth]{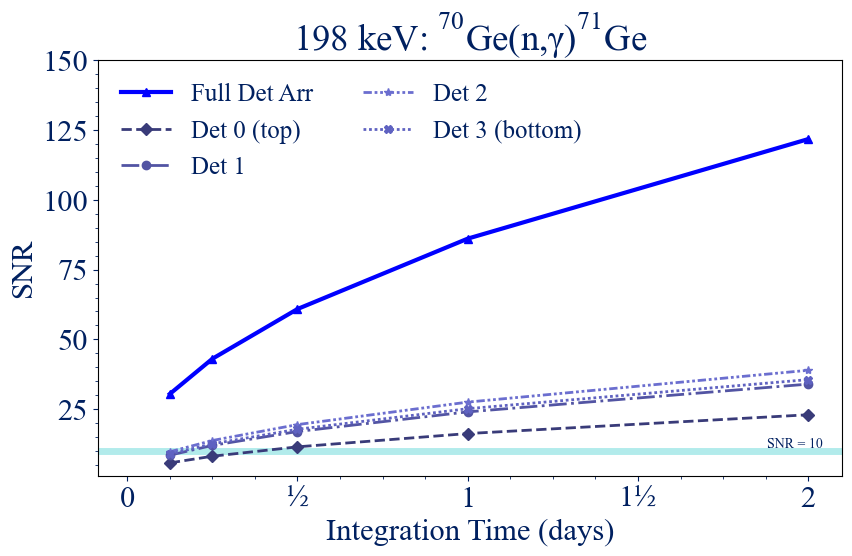}
    \includegraphics[width=0.49\linewidth]{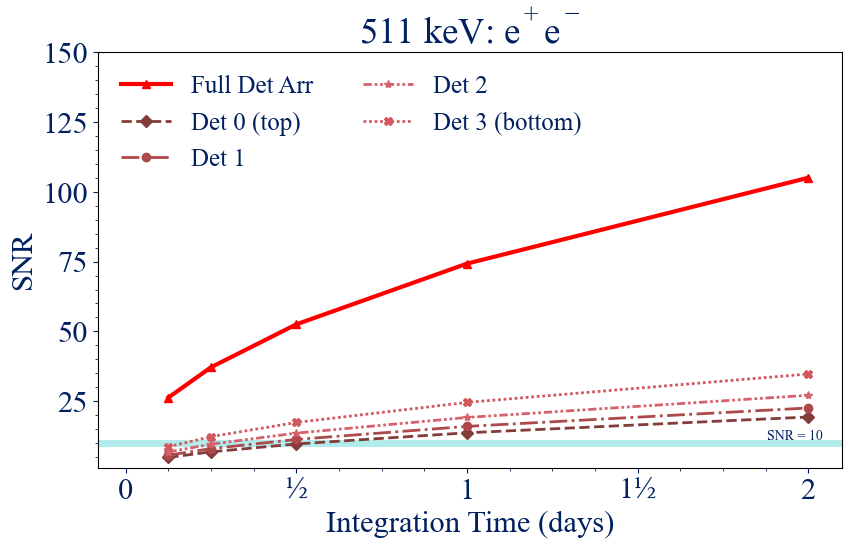}
    \includegraphics[width=0.49\linewidth]{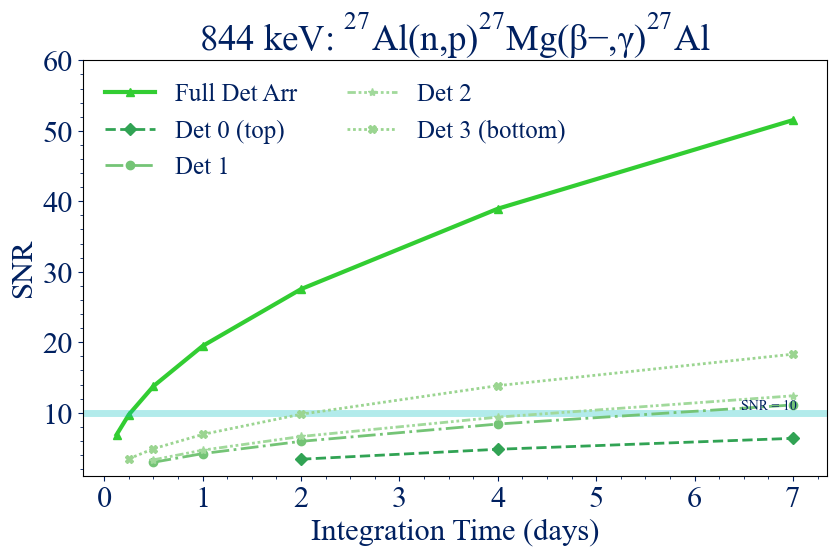}
    \includegraphics[width=0.49\linewidth]{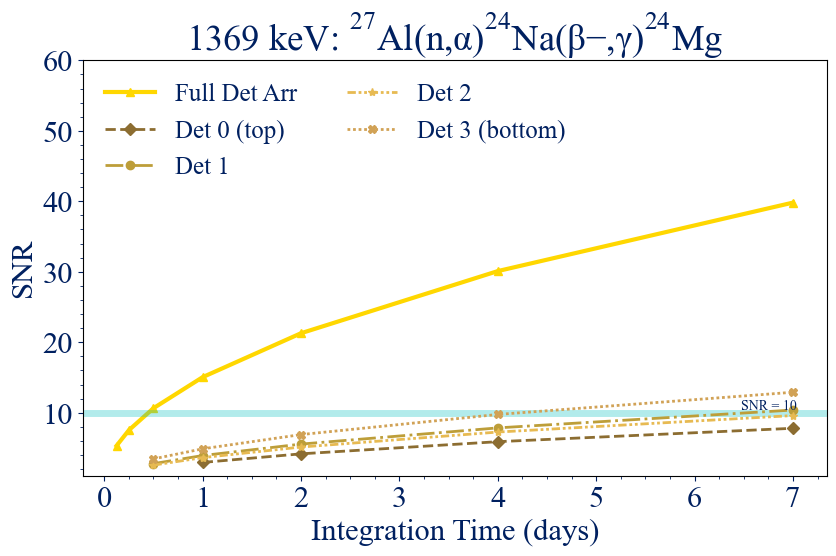}
    \caption{The evolution of the SNR value with time for various activation lines compared to individual GeDs is shown. The vertical dashed line in each plot marks the t\textsubscript{10} value for that line at the full instrument level. While the evolution is uniform for the 198 keV and 511 keV activation lines, the 844 keV and 1369 keV lines show strong positional dependence of the GeD, with Detector 3 attaining SNR=10 much earlier than the other detectors.}
    \label{fig:progressiveSNR}
\end{figure*}

\begin{figure}
    \centering
    \includegraphics[width=\linewidth]{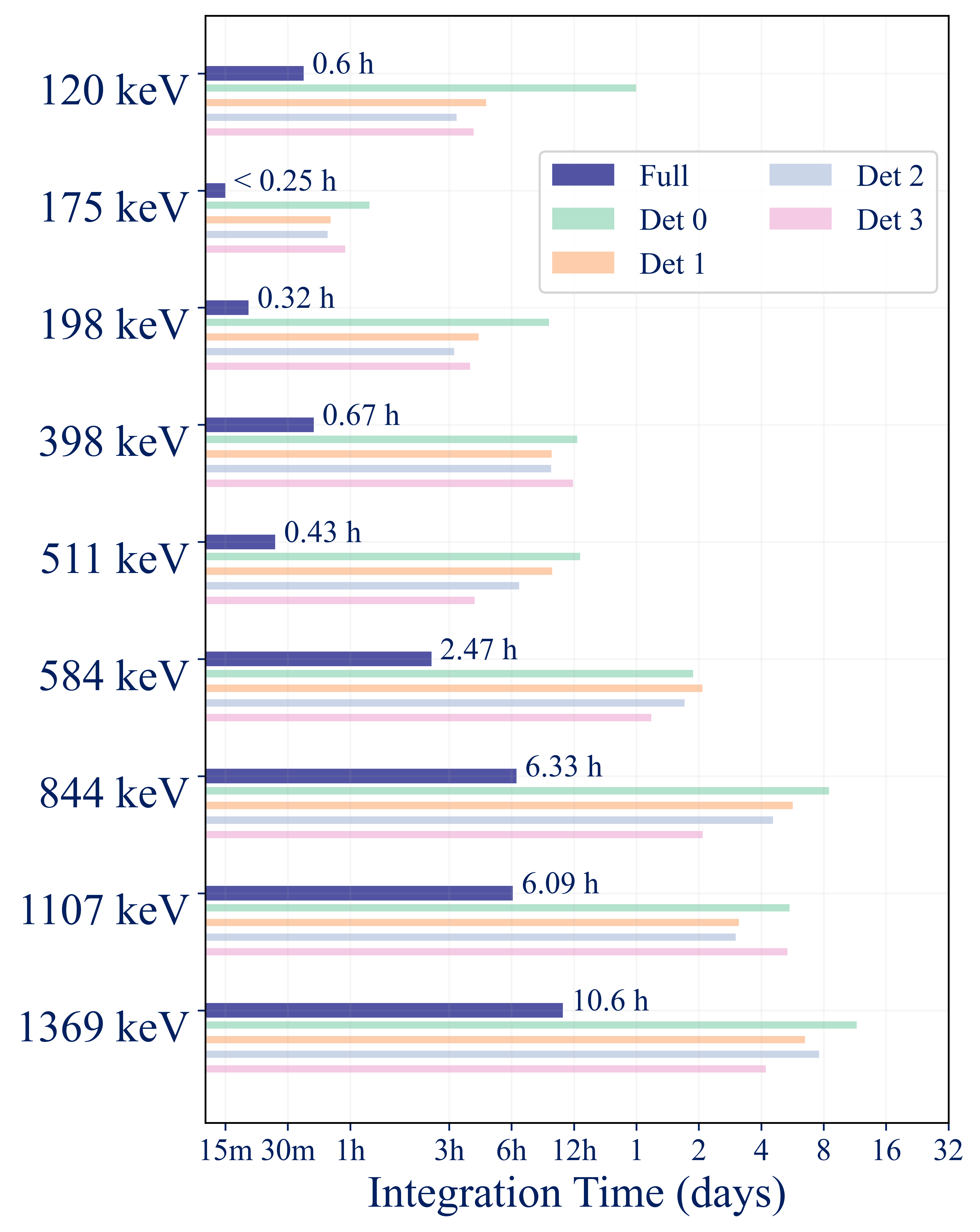}
    \caption{An illustration of the t\textsubscript{10} metric for nine activation lines that span from 100 keV to 1800 keV. The x-axis is cut off at 0.25 hours for clarity. The 844 keV and 1369 keV lines originate from the activation of aluminum in the cryostat, 511 keV is from the annihilation of positrons with multiple different sources, while the remaining lines are the result of GeD activation. }
    \label{fig:t10}
\end{figure}

In Fig. \ref{fig:modelfit}, we have shown the result of fitting the Gaussian+line model to an activation line at 1369 keV. The total flux uncertainty is on the order of 4\% and visually indistinguishable confirming that the fits are statistically robust. Fig. \ref{fig:linevsSNR} shows the progressive increase in the SNR value with longer integration times for various activation lines. The rate of increase follows $\text{SNR}\propto \sqrt t$, in accordance with Equation \ref{eq:SNR}. We calculated the integration time for which each curve passes the horizontal line at SNR=10, which we refer to as the t\textsubscript{10} metric. The figure demonstrates that ``stronger" activation lines have shorter t\textsubscript{10} values.

Similar plots are shown in Fig. \ref{fig:progressiveSNR} for the various detectors within the COSI $2\times2\times4$ detector array. The geometric mass model has an approximate, 4-fold rotational symmetry, and the count rates among the four detectors in a layer (shown in Fig. \ref{fig:sideview}) are uniform within 2\%. Here, Detector 0 refers to a GeD in the top-most layer (experiences the highest cosmic photon rate), while Detector 3 refers to a GeD in the bottom-most layer. 


\subsection{Time required for calibration}
The t\textsubscript{10} metric was calculated for the 53 activation lines, more than half of which were concentrated at energies below 500 keV. Of the 53 activation lines, 52 were observed to have a spectral width close to the instrument's resolution. By contrast, the 511 keV line was found to be made up of multiple beta decay components and highly kinematically broadened. \textit{Jean et al.} thus disfavored its use as a spectral calibration line \citep{jean2003spi}. Nevertheless, given its high count rate, this line might still be useful for gain calibration using a threshold SNR value greater than SNR=10. 

To identify prominent lines for spectral calibration, we selected the strongest lines, i.e., quickest to attain SNR=10, within consecutive 300 keV windows and spanning COSI's entire single-site energy range. This will allow us to monitor photopeak offsets at various energy scales as well as in the global gain characteristics. Based on the t\textsubscript{10} values, we have selected nine candidate lines for in-orbit calibration. The t\textsubscript{10} values at the instrument and individual GeD levels are shown in Fig. \ref{fig:t10}. 
Note that as the x-axis is in logarithmic scale, one can directly infer the percentage difference in t\textsubscript{10} values from the linear difference among bar lengths. 

Most components of the background spectrum (primary protons, albedo neutrons, etc.) are relatively constant with particle count rate variations on the order of 15\% over minute-long timescales \citep{gallego2025b}. However, the SAA component is strongly time-dependent, with short bursts of high background intensity when the satellite passes through the anomalous region. This results in brief periods of 10x higher particle irradiation. The effects of the SAA lasts throughout the orbit as some of the activated materials decay hours after the satellite exits the SAA region. In this study, we have used 7-day integrated data, which averages over these minute- and hour-long phenomena. Analyzing the temporal background variations caused by the SAA and the potential to use it for spectral calibration is the subject of future work.

\subsection{GeD position dependence of t$_{10}$ values}


In Figs. \ref{fig:progressiveSNR} and \ref{fig:t10}, we observe that most lines have the longest t\textsubscript{10} values for Detector 0. This can be attributed to the fact that the underlying continuum is highest in the top layer due to its direct exposure to photons from space. Thus, the continuum-to-line emission ratios can vary with the position of the GeD in the detector array. 

For Al-origin lines at 844 and 1369 keV, Detector 3 has by far the shortest t\textsubscript{10} values as most of the activation occurs in the thick base plate of the cryostat. 
For activation lines that originate in the GeD, such as 175 and 584 keV, the spatial distribution is approximately uniform as any of the 16 GeDs could be activated by extremely high energy particles. 

The 1107 keV GeD activation line deviates from these two trends wherein the middle detector layers take half the time as the top and bottom layers to reach SNR=10. This can be attributed to the higher penetration depth of the 1107 keV photons compared to lower energy photons. Photons created by activating one GeD are now more likely to be detected in a neighboring GeD. As the top and bottom layers have fewer GeDs surrounding them, they take longer to reach SNR=10.


\subsection{Parent process and material source}
\label{subsec:parent}

The parent processes can be identified by comparing the measured photopeak energies against a table of isotopes, and has already been done in previous gamma-ray telescope missions \citep{weidenspointner2003activationlines,wunderer2004activationmodelling,diehl2018integral}. 
It is especially important to determine the exact origins of a line emission to identify the exact energy of the line in order to utilize the line for spectral calibrations.

{In Table \ref{table:isotopes}}, we have summarized the activation line and their origins, including decay chains and half-lives. 
We have also noted the half-life of any intermediate step longer than the COSI-ACS coincidence time window ($t_c\sim1.5\ \mu$s, see Appendix A)  under ``other delayed activation lines". 

\section{Discussion}
\label{sec:discussion}
\subsection{Telemetry bandwidth and limitations}

Through this study, we have demonstrated that the instrument activation is sufficiently high to facilitate regular spectral monitoring. However, spectral calibration forms one of numerous calibration, health monitoring, and housekeeping tasks, which are cumulatively limited to 6 kbps of telemetry bandwidth (the bandwidth for multi-site interaction data is 84 kbps). Thus, while detector-level, spectral calibration can be done every 16 days in principle, the actual frequency of calibration may be modified depending on the gain fluctuation and radiation damage timescales. 


\subsection{Predictions for multi-site interaction data}
While this study has focused on characterizing delayed activation lines for in-orbit calibration, a few of these lines involve multi-step nuclear de-excitations. For example, the 198 keV line emission is in fact the sum of the 175 keV activation line and a 23 keV internal conversion electron. Similarly, the 1107 keV line emission can create a positron through pair production and result in two photons of 511 keV and 596 keV, also known as a single escape peak. These line emissions should therefore appear as individual peaks in the pre-reconstructed multi-site interaction spectrum and could also be used for spectral calibration. Moreover, the incident photon directions of these multi-site events can not be reconstructed as they arise from inherently distinct emission processes and will be flagged by the event reconstruction software \citep[Chapter 4]{zoglauer2005thesis}. Thus, in the future, we plan to extend this analysis to these multi-site events. {In the column labeled ``other delayed activation lines" in Table \ref{table:isotopes}}, we have listed potential multi-site event calibration lines from the nine candidate activation lines based on COSI's coincidence timescale and decay half-life. 

\begin{table*}[ht]
\centering
\caption{List of candidate activation lines for in-orbit calibration along with their decay chains and half-lives. The reactions in bold indicate the gamma-ray emitting step while bold-italics are used to denote radioactively unstable intermediate isotopes. EC stands for electron capture decay.}
\label{table:isotopes}
\begin{tabular}{llrr}
\hline
\hline
\\
\textbf{Activation Line} & \textbf{Decay Chain}                                                                                                                                           & \textbf{Half-life}                                                             & \textbf{Other Delayed Activation Lines}                                                          \\
\\
\hline
120 keV                  & 72Ge \textbf{(n, p$\gamma$)} 72Ga ($\beta^-$, $\gamma$) 72Ge                                                                                                  & 72*Ga 39.7ms                                                                              & 691 keV                                                                                                \\
\hline
175 keV                  & \multirow{2}{*}{\begin{tabular}[c]{@{}l@{}}70Ge \textbf{(n, $\gamma$)} \textit{71Ge} (EC, $\gamma$) 71Ga\\ 72Ge \textbf{(n, 2n $\gamma$)} \textit{71Ge} (EC, $\gamma$) 71Ga\end{tabular}} & \multirow{2}{*}{\begin{tabular}[c]{@{}r@{}}71*Ge 20ms\\ 71Ge 11d\end{tabular}} & \multirow{2}{*}{23 keV, 175 keV}                                                                 \\
198 keV                  &                                                                                                                                                                &                                                                                &                                                                                                  \\
\hline
398 keV                  & 70Ge \textbf{(n, 2n $\gamma$)} \textit{69Ge} (EC, $\gamma$) 69Ga                                                                                                        & \begin{tabular}[c]{@{}r@{}}69**Ge 2.8us \\ 69*Ge 5.1us\end{tabular}           & 87 keV, 311 keV                                                                                  \\
584 keV                  & \multirow{2}{*}{70Ge (n, 2n $\gamma$) \textit{69Ge} \textbf{(EC, $\gamma$)} 69Ga}                                                                                                & \multirow{2}{*}{69Ge 39h}                                                      & 574 keV                                                                                          \\
1107 keV                 &                                                                                                                                                                &                                                                                & \begin{tabular}[c]{@{}r@{}}596 keV, 607 keV, \\ 872 keV, 882 keV, 1117 keV\end{tabular} \\
\hline
844 keV                  & 27Al (n, p) \textit{27Mg} \textbf{($\beta^-$, $\gamma$)} 27Al                                                                                                                    & 27Mg 9m                                                                        & 1014 keV                                                                                         \\
1369 keV                 & 27Al (n, $\alpha$) \textit{24Na} \textbf{($\beta^-$, $\gamma$)} 24Mg                                                                                                             & \begin{tabular}[c]{@{}r@{}}24*Na 20ms\\ 24Na 15h\end{tabular}                  & \begin{tabular}[c]{@{}r@{}}472 keV, 858 keV\\ 1732 keV, 2243 keV, 2754 keV \end{tabular} \\  
\hline
511 keV                  & \textbf{e\textsuperscript{--} -- e\textsuperscript{+} annihilation} & - & - \\
\hline
\hline
\end{tabular}
\end{table*}

\subsection{Limitations and uncertainties in the background simulation model}\label{subsec:uncertainties}

The background simulation model and detector effects engine have a few known artifacts:
\begin{enumerate}
    \item The cosmic and albedo photon input models are cut-off below 100 keV. As these simulations were primarily performed for science studies within the 0.2--5 MeV energy range, 100 keV was determined as an optimal lower cut-off to save on CPU simulation time. Proper low-energy calibration is required to reconstruct photon direction from low-energy Compton scattering interactions and there are a few, strong activation lines of interest at energies below 100 keV as observed by previous telescopes such as RHESSI and INTEGRAL/SPI. 
    \item A few isotopes such as \textsuperscript{67}Ga (300, 393 keV) and \textsuperscript{69}Zn (438 keV) are under-produced compared to past INTEGRAL/SPI observations \citep{lonjou2005degradation} (also see COSI 2016 balloon observations in \citep[Figure 10]{gallego2025bottom}.) While a comprehensive comparison is yet to be performed, the observed differences in line strengths are believed to arise from variations in the internal secondary neutron spectra generated by COSI and INTEGRAL/SPI, which reflect the differences in their mass models. 
    \item Inelastic neutron scattering reactions such as \textsuperscript{72}Ge (n,n'$\gamma$) \textsuperscript{72}Ge might result in observations of a right-sided tail \citep[Figure 5.14]{boggs2002balloon, kierans2018thesis}. However, this is not currently captured by the detector effects engine, and we avoid using such lines for calibration.
    \item Fluorescent photons accompanying gamma-ray line emissions are often below our 18 keV electronics trigger threshold. For example, the 584 keV line is in fact the sum of a 574 keV line and a 10 keV fluorescent photon. This reaction should be more prominent at 574 keV as the 10 keV photon should not be recorded by the instrument. Such issues may stem from an incorrect modeling of the detector effects engine. 
    \item While we can extract the exact source of each GeD interaction from the Geant4 simulation logs, we did not log all our particle events during these simulations to save computational space. Simulation logs will help in precisely determining the list of particles created in each interaction, the parent process of each gamma ray photon, and enable distinguishing the various sources of error.
\end{enumerate}


As shown in Fig. \ref{fig:modelfit}, the statistical uncertainties in the Gaussian signal counts are on the order of $\Delta s / s \lesssim 4\%$ for a 7-day integration time. This translates to $\Delta t_\text{10} / t_\text{10} \lesssim 12\%$. These statistical uncertainties are much smaller than the multiple sources of systematic uncertainties in the background simulation, which range from inaccuracies in building the geometric mass model, to uncertainties in Geant4 nuclear reaction networks and capture cross-sections, modeling the various space radiation spectra, detector effects engine, and the orbital parameters of the spacecraft \citep{gallego2025b, gallego2025bottom,beechert2022calibrations, zoglauer2021cosi, sleator2019benchmarking}. 

At present, the simulation methods have been benchmarked against measurements from the COSI 2016 balloon flight \citep{kierans2017cosiballoon,gallego2025bottom}. Assuming our simulations are correct to an order of magnitude, a 10x lower background count rate will result in $\sqrt{10}\approx3.17$ times longer t\textsubscript{10} values, which will be slightly longer than the 36 day calibration timescale estimated in Section \ref{sec:analysis} at the individual detector level. A robust understanding of our background simulations is thus crucial, and we plan to provide a comprehensive analysis of the systematics and comparisons to past gamma-ray missions, such as RHESSI and INTEGRAL/SPI, in the future. We also plan to benchmark our background simulations using laboratory measurements of the activation spectrum under a simplified, proton radiation field.

\section{Conclusions}
\label{sec:conclusions}
Achieving COSI's science goals relies on our ability to maintain its spectral performance throughout the mission lifespan. In this work, we have employed Monte Carlo simulations of particle interactions performed using MEGAlib and developed a plan for in-orbit spectral calibration. \textit{Gallego et al.} \citep{gallego2025b} considered multiple background components ranging from cosmic rays, albedo particles, and trapped protons encountered during SAA passages to simulate a background spectrum for COSI. The simulations predict strong activation lines from germanium and aluminum, which can be used for in-orbit spectral monitoring and calibration. Additionally, our analysis predict that although COSI will encounter high levels of proton irradiation, the neutrons induced by these protons and subsequent neutron capture will be the dominant mode of activating spacecraft material.

We have calculated the strengths of 53 activation lines and selected nine candidate lines that span COSI’s entire energy range based on their t\textsubscript{10} values. By monitoring the photopeak energies of these line emissions, we have shown that the spectral performance of COSI’s detectors can be evaluated, and the effects of radiation damage and gain changes can be monitored at the instrument level every twelve hours. Activation line strengths at the individual detector level depend on their position in the detector array (top to bottom) and can be evaluated every 16 days. This is sufficiently frequent to correct photopeak shifts on the order of $\lesssim 0.05\%$. 

Although COSI’s background levels outside the SAA passages are predicted to be significantly lower than INTEGRAL/SPI and RHESSI due to differences in their orbits, these results rule out the need for an on board radioactive calibration source which would have increased the complexity of the spacecraft. 
These results also conform to COSI’s telemetry bandwidth for calibration and housekeeping data.


\section*{Appendix}
\subsection{COSI's Anti-Coincidence Subsystem (ACS)}
\label{subsec:BGO}
The COSI GeDs are supported by dense, high cross-section, bismuth germanate (BGO) shields on five out of the six faces, which stop stray photons from entering COSI's field of view \citep{ciabattoni2025benchmarking, patelgoals}. These shields are referred to as the anti-coincidence subsystem. Additionally, due to their high interaction cross-sections, they also efficiently shield laterally approaching high-energy particles. 

{However, high-energy particles that are only partially stopped by the BGO or enter through COSI's field of view can activate detector materials and result in gamma-ray decay photons. A GeD-BGO coincidence time window $t_c$ can be set to reject gamma-ray photons that are coincident with an incident high-energy particle. Based on the value of $t_c$, there are two types of gamma-ray emissions: ``Prompt" emission refers to those gamma rays arising from excited nuclei with half-lives $t_{1/2} \ll t_c$ and these are suppressed by the ACS \citep{jean2003spi, gallego2025b}.} 

On the other hand, gamma-ray emissions with $t_{1/2} \gg t_c$ at any upstream point in their decay chain are called ``delayed" emissions. These photons cannot be rejected by the ACS as they are emitted after the high-energy particle interacts with the BGO. Thus, the background gamma-ray spectrum that COSI will observe only contains this second category of delayed emissions. 
For a coincidence time value of $t_c \sim 1.5\ \mu$s, the prompt emissions vastly outnumber the delayed emissions and the ACS suppresses $>90\%$ of the background rate \citep{gallego2025c}. 


\section*{Acknowledgment}

This work was supported by the NASA Astrophysics Research and Analysis (APRA) program grant 80NSSC22K1881. SG acknowledges support from Deutsches Zentrum fur Luft-und Raumfahrt (DLR) grant 50OO2218. Resources supporting this work were provided by National High Performance Computing (NHR) South-West at Johannes Gutenberg University Mainz.
Resources supporting this work were provided by the NASA High-End Computing (HEC) Program through the NASA Advanced Supercomputing (NAS) Division at Ames Research Center. The Compton Spectrometer and Imager is a NASA Explorer project led by the University of California, Berkeley with funding from NASA, United States under contract 80GSFC21C0059.


\bibliographystyle{IEEEtran}
\bibliography{bibliography}

\end{document}